\batchmode
\documentclass[english,aps,preprint]{revtex4}
\usepackage[T1]{fontenc}
\usepackage[latin9]{inputenc}
\usepackage{bm}
\usepackage{amsmath}
\usepackage{amssymb}
\usepackage{esint}

\makeatletter

\DeclareRobustCommand{\greektext}{%
  \fontencoding{LGR}\selectfont\def\encodingdefault{LGR}}
\DeclareRobustCommand{\textgreek}[1]{\leavevmode{\greektext #1}}
\DeclareFontEncoding{LGR}{}{}
\DeclareTextSymbol{\~}{LGR}{126}
\newcommand{\lyxmathsym}[1]{\ifmmode\begingroup\def\b@ld{bold}
  \text{\ifx\math@version\b@ld\bfseries\fi#1}\endgroup\else#1\fi}

\@ifundefined{textcolor}{}
{%
 \definecolor{BLACK}{gray}{0}
 \definecolor{WHITE}{gray}{1}
 \definecolor{RED}{rgb}{1,0,0}
 \definecolor{GREEN}{rgb}{0,1,0}
 \definecolor{BLUE}{rgb}{0,0,1}
 \definecolor{CYAN}{cmyk}{1,0,0,0}
 \definecolor{MAGENTA}{cmyk}{0,1,0,0}
 \definecolor{YELLOW}{cmyk}{0,0,1,0}
}

\makeatother

\usepackage{babel}
\begin{document}

\title{THE FRACTIONAL LONDON EQUATION AND THE FRACTIONAL PIPPARD MODEL FOR
SUPERCONDUCTORS}

\author{José Weberzpil}

\email{josewebe@gmail.com}

\affiliation{$^{*}$Universidade Federal Rural do Rio de Janeiro - UFRRJ- Instituto
Multidisciplinar- IM- Departamento de Tecnologia e Linguagens- DTL-
Brazil}
\begin{abstract}
With the discovery of new superconductors there was a running to find
the justifications for the new properties found in these materials.
In order to describe these new effects some theories were adapted
and some others have been tried. In this work we present an application
of the fractional calculus to study the superconductor in the context
of London theory. Here we investigated the linear London equation
modified by fractional derivatives for non-differentiable functions,
instead of integer ones, in a coarse grained scenario. We apply the
fractional approach based in the modified Riemann-Liouville sense
to improve the model in order to include possible non-local interactions
and the media. It is argued that the effects of non-locality and long
memory, intrinsic to the formalism of the fractional calculus, are
relevant to achieving a satisfactory phenomenological description.
In order to compare the present results with the usual London theory,
we calculated the magnetic field distribution for a mesoscopic superconductor
system. Also,\textbf{ }a\textbf{ }fractional Pippard-like model is
proposed to take into account the non-locality beside effects of interactions
and the media. We propose that $\alpha$ parameter of fractionality
can than be used to create an alternative way to characterize superconductors.
\end{abstract}
\maketitle

\section{Introduction}

The non-integer order calculus, also called or fractional calculus
(FC) has become an promising emergent tool for scientific and applied
research. A landscape of various different fields have already been
touched by this powerful alternative for modeling. Thinking in terms
of complex systems, there exist studies that make uses of these for
modeling physical, biological, human and social systems, among other.
In the scope of several research fields, including physics, efforts
for the understanding connections between the dynamics and complex
systems have been found in the scientific literature, specially in
the non local theories. Among the motivations, one is that the use
of these theories may yield a much more elegant and effective treatment
of problems in science. The FC provides us with a set of mathematical
tools to generalize the concept of derivative and integral operators
with integer order to their respective extensions of an arbitrary
real order \cite{Herrmann}. Nonlocal theories and memory effects
have been thought as connected to complexity and admit a treatment
in terms of FC \cite{Cresus-Everton,Weberszpil-Aspects}. In this
context, the nondifferentiable nature of the microscopic dynamics
may be connected with time scales so as to approach questions in the
realm of complex systems \cite{Weberszpil}. 

In a coarse grained scenario, it was recently proposed a simple alternative
definition to the Riemann-Liouville derivative \cite{Jumarie,Jumarie2},
called modified Riemann-Liouville (MRL) fractional derivative, that
has the advantages of both the standard Riemann-Liouville and Caputo
fractional derivatives: it is defined for arbitrary continuous (non
differentiable) functions and the fractional derivative of a constant
is equal to zero. This kind of fractional calculus approach seems
to give a mathematical framework for dealing with dynamical systems
defined in coarse-grained spaces and with coarse-grained time and,
to this end, to use the fact that fractional calculus appears to be
intimately related to fractal and self-similar functions. We would
like to stress that the choice of MRL approach, besides the points
already mentioned, is justified by the fact that chain and Leibniz
rules acquires a simpler form, which helps a great deal if changes
of coordinates are performed. Moreover, causality seems to be more
easily obeyed in a field-theoretical construction if we adopt this
approach.

The non-differentiability and randomness are mutually related in their
nature, in such a way that studies in fractals on the one hand and
fractional Brownian motion on the other hand are often parallel \cite{Jumarie}.
A function which is continuous everywhere but not always differentiable
necessarily exhibits random-like or pseudo-random-features, in the
sense that various samplings of this functions on the same given interval
will be different. This may explain the huge amount of literature
which extends the theory of stochastic differential equation to stochastic
dynamics driven by fractional Brownian motion. The most natural and
direct way to question the classical framework of physics is to remark
that in the space of our real world, the generic point is not infinitely
small (or thin) but rather has a thickness. A coarse-grained space
is a space in which the generic point is not infinitely thin, but
rather has a thickness; and here this feature is modeled as a space
in which the generic increment is not $dx$, but rather $(dx)^{\alpha}$
and likewise for coarse grained with respect to the time variable
t.

In this paper we have worked out the linear London equation \cite{Gennes}
modified by fractional derivatives for non-differentiable functions
\cite{Jumarie,Jumarie2}, instead of integer ones, in a coarse grained
scenario. We also proposed a fractional Pippard model for superconductors.
We applied the fractional approach in the MRL sense to improve the
model in order to include possible non-local interactions and the
influence of the media. It is argued that the effects of non-locality
and long memory, intrinsic to the formalism of the fractional calculus,
are relevant to achieving a satisfactory phenomenological description.
By considering fractional derivatives in space, a generalized fractional
Laplacian is introduced and by means of a transformation of variables
called complex transform \cite{Fractional complex transform} we construct
an solve a fractional London equation. Also, based on a Chambers model
and following the Dressel and Gruner's book \cite{Dressel- Gruner-Book},
we proposed a fractional Pippard model for superconductors.

This paper is outlined as follows: In section II, we give some background
about the MRL formalism, In section III we develop the fractional
London equation with the solution. In section IV is devoted to review
some aspects of the Chambers development for Chambers formula. In
section V we present the fractional Pippard-like model based on Chambers
formula and in section V we cast our the concluding comments and prospects
for further investigation.

\section{Modified Riemann-Liouville Fractional Derivative}

The well-tested definitions for fractional derivatives, so called
Riemann-Liouville and Caputo have been frequently used for several
applications in scientific periodic journals. In spite of its usefulness
they have some dangerous pitfalls. Recently, it was proposed the MRL
definition for fractional derivative \cite{Jumarie,Jumarie2}, and
its basic definition is given by

\begin{eqnarray}
D^{\alpha}f(x) & = & {\displaystyle \lim_{x\rightarrow0}\, h^{-\alpha}\sum_{k=0}^{\infty}\,(-1)^{k}\,{\alpha \choose k}\, f\left(x+(\alpha\,-\, k)h\right)}=\nonumber \\
 & = & \frac{{1}}{\Gamma(1-\alpha)}\frac{d}{{dx}}\int_{0}^{x}(x\,-\, t)^{-\alpha}\left(f(t)-f(0)\right)dt;\nonumber \\
0 & < & \alpha<1.\label{eq:1}
\end{eqnarray}

Some advantages can be cited, first of all, using the MRL definition
we found that derivative of constant is zero, and second, we can use
it so much for differentiable as non differentiable functions. They
are cast as follows: 

(i) Simple rules:

\begin{eqnarray}
D^{\alpha}K & = & 0,\nonumber \\
Dx^{\gamma} & = & \frac{\Gamma(\gamma+1)}{\Gamma(\gamma+1-\alpha)}x^{\gamma-\alpha},\;\gamma>0,\nonumber \\
(u(x)v(x))^{(\alpha)} & = & u^{(\alpha)}(x)v(x)+u(x)v^{(\alpha)}(x).
\end{eqnarray}

(ii) Simple Chain Rules:

\begin{equation}
\frac{d^{\alpha}}{dx^{\alpha}}f[u(x)]=\frac{d^{\alpha}f}{du^{\alpha}}\,\left(\frac{du}{dx}\right)^{\alpha},\label{eq:Chainrule nondif func-1}
\end{equation}

for non differentiable functions and

\begin{equation}
\frac{d^{\alpha}}{dx^{\alpha}}f[u(x)]=\frac{df}{du}\,\frac{d^{\alpha}u}{dx^{\alpha}},\label{eq:chain rule space-time coarse-1}
\end{equation}

for coarse-grained space.

Details of the formalism can be found in the cited references and
references therein.

Now that we have set up these fundamental expressions, we are ready
to carry out the calculations of main interest, the construction and
the solutions to our fractional London equation and a development
of a fractional Pippard model.

\section{Linear Fractional London Equations}

For investigating the magnetic-fi{}eld distribution feature in the
fractional formalism, we need the modifi{}ed London equation. For
comparison with that in the integer case we will fi{}rst review the
derivation of the modifi{}ed London equation which is usually written
as \cite{Gennes}:

\begin{equation}
\bm{B}+\lambda^{2}\nabla\times(\nabla\times\bm{B})=0.
\end{equation}
where $\lambda$ is the London penetration depth and $\bm{B}$ is
the Magnetic induction vector field.

with the standard form \cite{Tinkham96},

\begin{equation}
\lambda^{2}\nabla^{2}\bm{B}(r)-\bm{B}(r)=-\Phi_{0}\delta^{(2)}(\mathbf{r})\hat{z},\label{eq:magnetic equation Tinkham}
\end{equation}
Here $\Phi_{0}=2\pi\hbar/e^{*}$ is the quantum flux and $\nabla^{2}$
is the vectorial Laplacian. The magnetic field is in $\hat{z}$ direction
and depends only on radial coordinates $r$.

This equation have a well known exact solution given by:
\begin{eqnarray}
B & = & \frac{\Phi_{0}}{2\pi\lambda}K_{0}(\frac{r}{\lambda}),
\end{eqnarray}

where $K_{0}$ is the zero order Hankel function.

With this 2-D geometry we can write the operator $\nabla^{2}$ as

\textbf{
\begin{equation}
\frac{1}{r}\frac{d}{dr}(r\frac{d}{dr}),
\end{equation}
}letting to a rewrite eq. \eqref{eq:magnetic equation Tinkham} as

\begin{equation}
-\frac{\lambda^{2}}{r}\frac{d}{dr}(r\frac{dB}{dr})+B(r)=\Phi_{0}\delta^{(2)}(\mathbf{r}-\mathbf{r}_{0}).\label{eq:London Integer}
\end{equation}

In order to rewrite the equation above with fractional derivatives
we use a generalized fractional Laplacian, in the MRL's sense, of
a form \cite{Fractional vector,Stillinger} given by

\begin{eqnarray}
\frac{1}{r^{\alpha}}D_{r}^{\alpha}(r^{\alpha}D_{r}^{\alpha}B(r)) &  & .
\end{eqnarray}
 with this fractional operator we can write the fractional London
equations as

\begin{eqnarray}
\frac{\lambda^{2\alpha}}{r^{\alpha}}D_{r}^{\alpha}(r^{\alpha}D_{r}^{\alpha}B(r)) & +B(r)=\Phi_{0}\delta^{(2)}(\mathbf{r}-\mathbf{r}_{0}) & .\label{eq:Fractional London}
\end{eqnarray}

Now, to solve eq. \eqref{eq:Fractional London} we apply a change
of variable called fractional complex transform \cite{Fractional complex transform}.
The fractional complex transform is used to convert fractional differential
equations with modified Riemann-Liouville fractional derivatives into
ordinary differential equations, so that all analytical methods devoted
to advanced calculus can be easily applied to fractional calculus.

To proceed, we define the new variable 

\begin{equation}
R=\frac{pr^{\alpha}}{\Gamma(1+\alpha)}.
\end{equation}

With the help of the above transformation and the rule given by \ref{eq:chain rule space-time coarse-1},
we obtain an integer one of the similar form of eq.\eqref{eq:London Integer}
but with argument depending on variable$R$. and with known solution.
The final form of solution of the eq. \eqref{eq:Fractional London}
is given by

\begin{eqnarray}
B(r/\lambda)=\frac{\Phi_{0}}{2\pi\lambda^{\alpha}}K_{0}\left[\frac{\left(r/\lambda\right)^{\alpha}}{\Gamma(\alpha+1)}\right]
\end{eqnarray}

The above expression can be used to compare the behavior of this solution
to the well known integer one by the use of a algebraic or graphic
computational software. Some comments will be given in the concluding
remarks.

Local electrodynamics assumes that the response at one point in the
material only depends on the electric field at this point. This assumption
breaks down in the case of metals with a long mean free path and in
the case of superconductors with a long coherence length $\lyxmathsym{\textgreek{x}}_{0}$.
The anomalous skin effects become important if $\lyxmathsym{\textgreek{d}}<l$.
For superconductors the London limit is exceeded if $\lyxmathsym{\textgreek{l}}<\lyxmathsym{\textgreek{x}}_{0}$.
In both cases the influence of electrons, which feel a different electric
field at some distant point, becomes important. To take into account
the non-local effects beside the interactions and media measured by
a fractionality parameter, we propose a fractional Pippard-like model.

\section{The Chambers Formula and the Pippard's Model}

To work out a Pippard-like fractional model, we proceed as in \cite{Dressel- Gruner-Book}
and we give here the details for clarity in the procedure. The Chamber
model is used in the theory anomalous skin effect for the treatment
of the non-local conductivity. The anomalous skin effect is the name
given to the behavior of a metal under such conditions of purity and
low temperature that the high-frequency oscillations of electric field
and current are confined within a surface layer of thickness much
less than the mean free path \cite{Pippard-Skin effect}. I In this
model, the current density in a normal metal in which the electric
field varies over a mean free path is determined. For this, consider
the $\vec{q}$ moment dependent response is intimately related to
the non-local conduction where the current density $\vec{J}$ at the
position $\vec{r}$ is determined also by fields at other locations
$\vec{r}\neq\vec{r'}$. Following, an approximate expression for the
current density which depends on the spatial distribution of the applied
electric field is developed for a a situation that may occur in the
case of clean metals at low temperatures when the mean free path is
large. Considering in the sequence an electron at $\vec{r}$ and subjected
to an electric field $\vec{E}(\vec{r})$, moving to another position
taken to be the origin of the coordinate system, where the field is
different from that at initial position. However, because of collisions
with the lattice or impurities, the momentum acquired by the electron
from the field at $\vec{r}$ decays exponentially as the origin is
approached. This last consideration will be relaxed into the fractional
approach. The characteristic decay length defines the mean free path,
and the currents at the origin are the result of the fields $\vec{E}(\vec{r})$
within the radius of $l=v_{F}\lyxmathsym{\textgreek{t}}$ . The non-local
response follows the argumentation that when an electron moves from
a position $(\vec{r}-d\vec{r})$ to $\vec{r}$, it is influenced by
an effective field $\vec{E}(\vec{r})exp(-r/l)$ for a time $d\vec{r}/v_{F}$.
The momentum acquired $d\vec{p}(0)$ in the direction of motion and
at initial position is

\begin{equation}
d\vec{p}(0)=\frac{ed\vec{r}}{v_{F}}\frac{\vec{r}}{r}\cdot\vec{E}(\vec{r})exp(-r/l);\label{eq:dp}
\end{equation}
By integrating the eq. \eqref{eq:dp} from the origin to infinity,
the total change in momentum for an electron at the origin is found.
Performing this calculation for all directions allows to map out the
momentum surface in a non-uniform field, and the deviations from a
sphere centered at the origin constitute a current density $\vec{J}$.
The following arguments lead to the expression of this current density.
Only electrons residing in regions of momentum space not normally
occupied when the applied field is zero contribute to the current
density.

The density of electrons $\triangle N$ moving in a solid angle $d\Omega$
and occupying the net displaced volume in momentum space $\triangle P$
is

\begin{equation}
\triangle N=\frac{\triangle P}{P}N=\frac{(mv_{F})^{2}d\Omega dp}{\frac{4}{3}\pi(mv_{F})^{3}}N=\frac{3Nd\Omega dp}{4\pi v_{F}m},
\end{equation}
 where $P$ is the total momentum space volume. The contribution to
the current density from these electrons is

\begin{equation}
d\vec{J}=-\triangle N\, e\, v_{F}\frac{\vec{r}}{r}=-\frac{3Ne}{4\pi m}\frac{\vec{r}}{r}d\Omega dp.
\end{equation}

Substituting eq. \eqref{eq:dp} into this equation and integrating
over the currents given above yields 
\begin{equation}
\vec{J}(\vec{r}=0)=\frac{3\sigma_{dc}}{4\pi l}\int\frac{\vec{r}\left[\vec{r\cdot}\vec{E}(r)exp(-r/l)\right]}{r^{4}}d\vec{r},\label{eq:J-Chambers}
\end{equation}
since a volume element in real space is $r^{2}drd\Omega$ and $\sigma_{dc}=Ne^{2}\lyxmathsym{\textgreek{t}}/m=Ne^{2}l/(mv_{F})$.
Equation \eqref{eq:J-Chambers} represents the non-local generalization
of Ohm\textquoteright{}s law for free electrons, and reduces to $\vec{J}=\lyxmathsym{\textgreek{sv}}_{dc}\vec{E}$
for the special case where $l\rightarrow0$, as expected. The Chambers
formula is valid for finite momentum, but as the Fermi momentum is
not explicitly included its use is restricted to $\vec{q}<\vec{k_{F}}$,
and in general to the small $\vec{q}$ limit.

Note that, in view of $\lyxmathsym{\textgreek{sv}}_{dc}=Ne^{2}\lyxmathsym{\textgreek{t}}/m=Ne^{2}v_{l}l/m$,
the mean free path drops out from the factor in front of the integral.
For superconductors, the vector potential $\vec{A}$ is proportional
to $\vec{J}$, and with $\vec{E}=i(\lyxmathsym{\textgreek{w}}/c)\vec{A}$
we can write accordingly 
\begin{equation}
\vec{J}(0)\propto\int\frac{\vec{r}\left[\vec{r\cdot}\vec{A}(r,\omega)\right]}{r^{4}}F(r)d\vec{r}.
\end{equation}

With this background, we can now proceed to the obtainment of the
fractional Pippard-like model.

\section{The Fractional Pippard-Like Model}

Here, we will assume that the kernel function $F(r)$ have to take
into account that the charge carriers can be thought as pseudo-particles
that carries the implicit information of an effective field and the
media, attributing to each location a probability $\frac{(\frac{r-r'}{\xi})^{\alpha-1}}{\Gamma(\alpha)}.$
Another possibility is to attribute to $F(r)$ an stretched exponential,
that is, the two parameters Mittag-Leffler function as the probability
factor that substitute the exponential in the Pippard's original formula. 

The formula is than

\begin{equation}
\vec{J}(\vec{r})\propto\int\frac{\vec{r}\left[\vec{r\cdot}\vec{A}(r,\omega)\right]}{r^{4}}F(r)E_{\alpha\beta}(-r/l)d\vec{r}\label{eq:J sem subst Fr}
\end{equation}
 where $F(r)=\frac{(\frac{r-r'}{\xi})^{\alpha-1}}{\Gamma(\alpha)},$
$\xi$ is the correlation length and the two parameter Mittag-Leffler
functions is given by $E_{\alpha\beta}(x)=\underset{k=0}{\sum^{\infty}}\frac{x^{k}}{\Gamma(\alpha x+\beta)}$.
With the $F(r)$, eq. \eqref{eq:J sem subst Fr} is now 

\begin{equation}
\vec{J}(\vec{r}')\propto\frac{1}{\Gamma(\alpha)}\int(r-r')^{\alpha-1}K(A,r,\xi)d\vec{r}=I^{\alpha}K(A,r,\xi)\label{eq:Jfrac 1}
\end{equation}
 where 
\begin{equation}
K(A,r,\xi)=(\frac{1}{\xi})^{\alpha-1}\frac{\vec{r}\left[\vec{r\cdot}\vec{A}(r,\omega)\right]}{r{}^{4}}E_{\alpha\beta}(-r/l),
\end{equation}
and $I^{\alpha}K(A,r,\xi)$is the fractional integral of kernel $\mbox{\ensuremath{K(A,r,\xi).}}$

The fractional approach here is still justified by the argumentation
that the particle described by this formalism can be thought as an
pseudo-particle , as already commented, it would carry the information
of the media and the kind of interactions implicit in the equation
that describes his evolution. This pseudo-particle would be then ``dressed''
with information about media and interactions, and the solutions of
the fractional equation are, like the Green functions in condensed
matter physics, carrying additional information about iterations and
the media. Then, even if the media is not fractal due to not so high
energy regime, the fractional approach still makes sense to describe
the evolutions of a pseudo-particle. This means that, in the fractional
approach context, the particle interacts not as an isolated particle
but as a pseudo-particle.

The phenomenology here is in some sence similar to that of an anomalous
transport with different relaxation times, as in some non-Newtonian
viscous systems, but with anomalous correlation, with some ``memory
effect'' or heredity, that gives to each location a probability proportional
to a power of distance from the source, it leads to non-local fractional
effective theory. The $\alpha$ parameter of fractionality can than
be used to create an alternative way to characterize superconductors.

\section{Concluding Remarks}

In this work, by taking into account a non-differentiable space (coarse-grained),
we have obtained a fractional linear London equation in terms of a
fractional Laplacian with a sequential form of modified Riemann-Liouville
fractional derivatives. We claim that the novelty of our work is the
particular choice of formalisms with sequential modified fractional
derivatives and an adequate chain rules, applied in the superconductivity
area, leading to technique that creates a perspective to obtain solutions
for other similar problems like Preisach hysteresis model \cite{Supercond Hysteresis-Preisach,Dynam Preisach}.
Our solutions are worked out by means of a complex transform in the
space variable that permits to change the fractional London Equation
to an integer one with known Bessel like solutions. The MRL approach
of fractional calculus seems to be more adequate to deal with problems
that involve transformations in coordinates, since the chain and Leibniz
rules are less complicated. Since we are choosing to work with non-differentiable
coarse-grained space, no use of distributional generalized functions
or fractional powers of operators, neither the maintenance of semi-group
properties of exponents in the derivatives is made. 

Our solution of fractional linear London model indicates that by means
of a Meissner effect, the exclusion of the magnetic field is more
effective in the vicinity of the boarder of a vortex than in the integer
case but the residual field inside the vortex is greater than in the
integer case one. This can also indicates that the fractional model
could take into account influence of the contours of the grains to
the characteristics of the Meissner effect in a Type II superconductor
by means of a fractionality parameter $\alpha$.

The results agrees with standard integer order in the convenient limit
of $\alpha=1$.

Also, based on a Chambers model for the anomalous skin effect, we
have proposed a fractional Pippard-like model with non-local effects
that can gives rises to an effective theory for superconductors and
can be used to characterize and distinguish superconductors properties
by the measurement of a fractionality parameter.
\begin{acknowledgments}
The author wishes to express his gratitude to Prof. Isaias G. de Oliveira
(DEFIS-UFRRJ) by the discussions.

I am also very grateful to prof. Cresus F. L. Godinho (DEFIS-UFRRJ)
by the text revision.\end{acknowledgments}


\begin{thebibliography}{References}
\bibitem{Herrmann}Richard Herrmann, Common aspects of q-deformed
Lie algebras and fractional calculus, Physica A 389 4613-4622 (2010);
arXiv:1007.1084v1 {[}physics.gen-ph{]}. 

\bibitem{Cresus-Everton}E.M.C Abreu and C.F.L. Godinho, ''Fractional
Dirac bracket and quantization for constrained systems'', Phys. Rev.
E84 026608 (2011).

\bibitem{Weberszpil-Aspects}J. Weberszpil, C. F. L. Godinho, A. Cherman,
J. A. Helayël-Neto, ''Aspects of the Coarse-Grained-Based Approach
to a Low-Relativistic Fractional Schrödinger Equation'', arXiv:1206.2513v3
{[}math-ph{]} (2012).

\bibitem{Weberszpil}Cresus F.L. Godinho, J. Weberszpil, J.A. Helayël-Neto,
''Extending the D'Alembert Solution to Space-Time Modified Riemann-Liouville
Fractional Wave Equations'', Chaos, Solitons \& Fractals 45 765\textendash{}771
(2012).

\bibitem{Gennes}P.G. de Gennes, Superconductivity of Metal and Alloys,
(New York: Benjamin), 1999.

\bibitem{Jumarie}Guy Jumarie, ''\textit{Fractional Partial Differential
Equations and Modified Riemann-Liouville Derivative New Methods for
Solution}'', J. Appl. Math. \& Computing Vol. \textbf{24}, No. 1 -
2, pp. 31 - 48 (2007).

\bibitem{Jumarie2}G. Jumarie, ''Table of some basic fractional calculus
formulae derived from a modiffed Riemann-Liouville derivative for
non-differentiable functions'', Applied Mathematics Letters 22, 378-385
(2009).

\bibitem{Fractional complex transform}Li ZB, He JH. ''Fractional
Complex Transform for Fractional Differential Equations'', Mathematical
and Computational Applications, \textbf{15} (5) 970-973 (2010).

\bibitem{Fractional vector}Ming-Fan Li, Ji-Rong Ren, Tao Zhu, ''Fractional
Vector Calculus and Fractional Special Function'', arXiv:1001.2889v1
{[}math-ph{]}(January 2010).

\bibitem{Stillinger}Frank H. Stillinger, ''Axiomatic basis for spaces
with non integer dimension'', Journal of Mathematical Physics, Vol.
\textbf{18}, No.6 (1977).

\bibitem{Tinkham96}Michael Tinkham, Introduction to superconductivity,
McGraw-Hill Science Engineering Math (1995); eq. 5.13.

\bibitem{Dressel- Gruner-Book}Dressel M., Gruner G. - Electrodynamics
of Solids. Optical Properties of Electrons in Matter, Cambridge University
Press (2002).

\bibitem{Pippard-Skin effect}A. B. Pippard, ''The Anomalous Skin
Effect in Anisotropic Metals'', Proceedings of the Royal Society of
London. Series A, Mathematical and Physical Sciences, \textbf{224},
No. 1157, 273-282 (1954).

\bibitem{Supercond Hysteresis-Preisach}I. D. Mayergoyz and T. A.
Keim, ''Superconducting hysteresis and the Preisach model'', J. Appl.
Phys. 67, 5466 (1990).

\bibitem{Dynam Preisach}Miklós Kuczmann, ''Dynamic Preisach hysteresis
model'', Journal of Advanced Research in Physics 1(1), 011003 (2010).\end{thebibliography}
\end{document}